\documentstyle[aps,prl,multicol,epsfig]{revtex}   
\tighten

\begin{document}

\title{Josephson nanocircuit in the presence of linear quantum noise}

\author{E. Paladino$^{(1,2,3)}$, F. Taddei$^{(1,2,3)}$, 
G. Giaquinta $^{(1,2,3)}$, and G. Falci
$^{(1,2)}$
\footnote{Corresponding author: G. Falci. 
Present address: DMFCI, Universit\'a di Catania, Viale A. Doria 6,
95125 Catania, Italy. E-mail: gfalci@dmfci.unict.it}
}

\address{
$^{(1)}$ Dipartimento di Metodologie Fisiche e Chimiche (DMFCI),
        Universit\`a di Catania,\\
        viale A. Doria 6, 95125 Catania, Italy\\
$^{(2)}$ INFM, Unit\'a di Catania, Catania, Italy \\
$^{(3)}$ NEST-INMF, Pisa, Italy  }

\date{\today}

\maketitle

\begin{abstract}
We derive the effective hamiltonian for a charge-Josephson qubit in 
a circuit with no use of phenomenological arguments, showing 
how energy renormalizations induced 
by the environment appear with no need of 
phenomenological counterterms. This analysis may be important for 
multiqubit systems and geometric quantum computation.
\end{abstract}

\begin{small}
\end{small}

\begin{multicols}{2}
\narrowtext  

\begin{figure}[btp]
%h=here, t=top, b=bottom, p=separate figure page
\begin{center}\leavevmode
\includegraphics[width=0.8\linewidth]{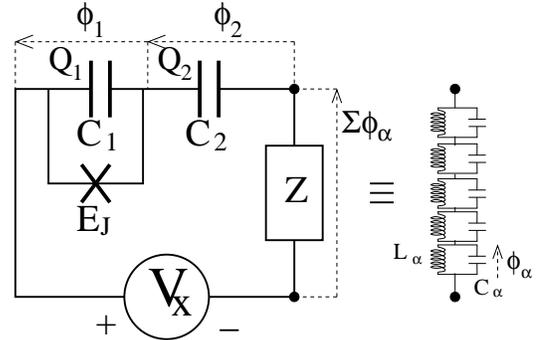}
\caption{The Cooper pair box in an electromagnetic environment. Fluctuations
are due to the impedence $Z(\omega)$, which is modeled by a suitable 
infinite LC transmission line
}\label{fig:circuit}\end{center}\end{figure}

Josephson junction based nanocircuits
have been proposed for the implementation 
of quantum gates\cite{kn:rmp} and quantum coherent behavior has been 
recently observed\cite{kn:exp} in charge-based devices 
(charging energy $E_C$ larger than
the Josephson energy $E_J$). Decoherence in these devices is due to 
several sources\cite{kn:rmp}, as fluctuations of the circuit, 
backaction of the measuring apparatus or noise due to background charges 
in the substrate\cite{kn:paladino}. Fluctuations 
of the circuit are modeled by coupling the system to an environment of 
harmonic oscillators\cite{kn:dissipative} which mimics the external 
impedences (see Fig.\ref{fig:circuit}). 
An effective hamiltonian $H_{eff}$ is considered, which represents 
a spin-boson model (or a multistate version), the central variable being 
the charge $Q$ in the island and 
the environment being fixed in a phenomenological way\cite{kn:rmp}. 
The environment produces decoherence and energy
shifts, which may in principle be large. In dissipative quantum 
mechanics shifts are usually treated either by introducing 
counterterms\cite{kn:dissipative} or by writing $H_{eff}$ 
in terms of renormalized quantities\cite{kn:dissipative}. We present here 
a model for the electromagnetic environment and we derive a multistate
$H_{eff}$ using no phenomenological 
argument. This has two motivations. First in 
principle bare circuit parameters are well defined and tunable, so we want 
to know precisely how this reflects on $H$.
Second the role of induced shifts, which is minor in 
the devices of Refs.\cite{kn:exp}, may be crucial 
in various situations (e.g. geometric quantum computation\cite{kn:nature}, 
dynamics of registers 
and error correction devices).

We consider the Cooper pair box\cite{kn:rmp} of Fig.\ref{fig:circuit}. 
The external impedance is modeled by a suitable $LC$ transmission line
and the Lagrangian of the system is 
$${L} \;=\; \sum_{i=1,2} {C_i \dot{\phi}_i^2 \over 2} 
	- V_J(\phi_1) 
+ 
	\sum_\alpha \bigl[ 
	{C_\alpha \dot{\phi}_\alpha^2 \over 2} - 
	{\phi_\alpha^2 \over 2 L_\alpha}
	\bigr] 
$$
where $\dot{\phi}$ are voltage drops and the Josephson energy is
$V_J(\phi_1) =	- E_J  \cos (2 e \phi_1/ \hbar) $.
The environment is fully specified by the elements $C_\alpha$ and 
$L_\alpha$. The circuit is introduced by the constraint
$\dot{\phi}_1 + \dot{\phi}_2 + 	\sum_\alpha 
\dot{\phi}_\alpha = V_x$, which allows to eliminate one variable and
to write
\begin{eqnarray}
\label{eq:superconducting-box-dissipativa-lagrangian-2}
{L} &=& {C_\Sigma\, \dot{\eta}^2 \over 2} - 
	V_J\bigl(2 e (\kappa_2 \Phi -  \eta)/ \hbar \bigr)
	  + L_b
\\
 L_b &=&{1 \over 2}\, C_e\, \dot{\Phi}^2 \,+ \,
	\sum_\alpha \bigl[ 
	{C_\alpha \over 2} \; \dot{\phi}_\alpha^2 \,- \, 
	{1 \over 2 L_\alpha} \; \phi_\alpha^2
	\bigr]
\nonumber
\end{eqnarray}
where $\eta =  \kappa_2 \phi_2 -  \kappa_1 \phi_1$, 
$C_{\Sigma}= \sum_{i} C_{i}$, $C_e=C_{1}C_{2}/C_{\Sigma}$, 
$\kappa_{1,2} = C_{1,2}/C_{\Sigma}$, and 
$\dot{\Phi} = V_x - 	\sum_\alpha  \dot{\phi}_\alpha$.

We next diagonalize $L_b$ and obtain the form
$$
{ L}_b  \;=\; 
\sum_\alpha \Bigl[ 
	{m_\alpha \over 2} \; \dot{x}_\alpha^2 \,- \, 
	{m_\alpha \omega^2_\alpha \over 2} \; x_\alpha^2
	\Bigr]\,- \, C_e V_x \, \sum_\alpha  d_\alpha \dot{x}_\alpha \, .
$$

In order to determine the parameters, notice that
$L_b$ describes a series $C_e-Z$ circuit, with 
$\sum_\beta  d_\beta \dot{x}_\beta = \sum_\alpha  \dot{\phi}_\alpha = V_Z$. 
By comparing the linear response with the known classical dynamics of $V_Z$ 
(as explained in Refs.\cite{kn:dissipative}) we determine the spectral 
density ($\omega > 0$)
$$
J^\prime(\omega)
= \sum_{\alpha} \;  { \pi d^2_{\alpha} \delta(\omega - \omega_\alpha)
\over 2 m_\alpha \omega_\alpha}
=
\mathrm{Re} \Bigl[ {Z(\omega) / \omega  \over 1 + i \omega Z(\omega) C_e}
\Bigr]
$$
We stress that $L_b$ is quadratic therefore the procedure above is an exact 
way to perform the diagonalization, which uses classical circuit theory
as a tool. 
It's validity for quantum harmonic oscillators is guaranteed by the 
Ehrenfest theorem. 

To get rid of the $\Phi$ in the potential term in 
Eq.(\ref{eq:superconducting-box-dissipativa-lagrangian-2}), 
we perform a (canonical) transformation 
$\chi = \eta - \kappa_2 \Phi$ and obtain the total Lagrangian
$L= L_a + L_b$ where
$$
{L}_a = {C_\Sigma \over 2}
	\bigl( \dot{\chi} + \kappa_2  V_x - 
	\kappa_2 \sum_\alpha  d_\alpha \dot{x}_\alpha
	\bigr)^2 
	+ 
	V_J( 2 e \chi/ \hbar)
$$
One can verify that the variable canonically conjugated to $\chi$ 
is the charge $Q$ in the island. The Hamiltonian corresponding to $L$ reads
\begin{eqnarray}
{H} 
&=&
{Q^2 \over 2 C_1} \; \,+\, Q \; \kappa_2
\sum_\alpha   { d_\alpha \over m_\alpha}  \; p_\alpha 
\,-\, E_J \, \cos \Bigl( {2 e \over \hbar} \chi \Bigr) 
\nonumber\\ && \hskip10pt  +   
\sum_\alpha \Bigl[ {p_\alpha^2 \over 2 m_\alpha} \,+\, 
 {1 \over 2} \, \omega^2_\alpha \, x_\alpha^2 \Bigr] 
 \,+\,  C_e V_x \; \sum_\alpha {d_\alpha  \over m_\alpha} \; p_\alpha
\nonumber
\end{eqnarray}
where $p_\alpha$ are conjugated to $x_\alpha$ and we used the relation
$\sum_\alpha d_\alpha^2 / m_\alpha = 1/C_e$. The system variable $Q$ is coupled
with the momenta $p_\alpha$ of the environment.

A further canonical 
transformation of the environment 
($
\tilde{x}_\alpha = {p_\alpha/( m_\alpha \omega_\alpha)}$,  
$\tilde{p}_\alpha = - m_\alpha \omega_\alpha  x_\alpha$) yields a hamiltonian
where $Q$ is coupled
with the coordinates
\begin{eqnarray}
\label{eq:ZC-superc-box-hamiltonian-3}
{H}_{eff} &=& {Q^2 \over 2 C_1} + V_J\bigl( {2 e \over \hbar} \chi \bigr) +
 \kappa_2 Q \sum_\alpha c_\alpha \, \tilde{x}_\alpha 
\nonumber\\ &+&
\sum_\alpha \Bigl[ 
	{\tilde{p}_\alpha^2 \over 2 m_\alpha} \,+ \, 
	{m_\alpha \omega^2_\alpha \over 2} \; \tilde{x}_\alpha^2
	\Bigr]\,+ \, C_e V_x \, 
\sum_\alpha c_\alpha \, \tilde{x}_\alpha
\nonumber
\end{eqnarray}
where
$c_\alpha = {d_\alpha \omega_\alpha}$ and we introduce the spectral density
$J(\omega)
= \pi  \sum_{\alpha} \delta(\omega - \omega_\alpha) c^2_{\alpha}/ 
(2 m_\alpha \omega_\alpha) = \omega^2 \,J^\prime(\omega)$. 

If we isolate the dc bias  $V_x(t) =V_x+ \delta V_x(t)$
and redefine 
$\tilde{x}_\alpha + c_\alpha C_e V_x/(m_\alpha \omega^2_\alpha) 
\, \to \, x_\alpha$
we finally obtain
%\vspace{-12pt}
\begin{eqnarray}
\label{eq:ZC-superc-box-hamiltonian-4}
{H}_{eff} &=& {Q^2 \over 2 C_1} - \kappa_2 V_x  Q + 
V_J\bigl( {2e \over \hbar} \chi \bigr)
+ Q  \kappa_2 \sum_\alpha c_\alpha  x_\alpha 
\nonumber\\
&+&
\sum_\alpha \Bigl[ 
	{p_\alpha^2 \over 2 m_\alpha} +  
	{m_\alpha \omega^2_\alpha \over 2} x_\alpha^2
	\Bigr] +  C_e \delta V_x(t)  
\sum_\alpha c_\alpha  x_\alpha \quad
\nonumber
\end{eqnarray}
Notice that the capacitance $C_1$ (and not $C_\Sigma$) enters the $Q^2$ term 
and if we put $c_\alpha=0$ we do not obtain the Cooper pair box 
hamiltonian. This is correct because the 
environment represents global fluctuations of the circuit, not only of $Z$.
Notice that a static $Q$ shifts
the equilibrium positions of the oscillators and also produces a $Q$-dependent 
shift of the zeroes of their energies. We can single out the corresponding 
term and focus on it. 
In the situations described in 
Refs.\cite{kn:dissipative} this term should   
be canceled by introducing a counterterm, because we have informations only 
about renormalized effective parameters of the model. In our case we have 
information about the bare parameters therefore there is no reason to cancel 
the $Q$-dependent shift of the zero of the oscillator energies. It 
can be reabsorbed 
in the charging energy if we write the oscillator hamiltonian using the 
shifted values $x_\alpha + c_\alpha \kappa_2  Q / (m_\alpha \omega^2)$, which
produces the extra term 
$- \kappa_2^2  Q^2 \sum_\alpha {c_\alpha^2 / (2  m_\alpha \omega_\alpha^2)}
= - Q^2/(2C_e)$ and 
\begin{eqnarray}
\label{eq:ZC-superc-box-hamiltonian-5}
{ H}_{eff} &=& {Q^2 \over 2 C_\Sigma} - \kappa_2 V_x  Q + 
V_J\bigl( {2e \chi \over \hbar}\bigr) 	+  C_e \delta V_x%(t)  
\sum_\alpha c_\alpha  x_\alpha
\nonumber\\&+&
\sum_\alpha \Bigl[ 
	{p_\alpha^2 \over 2 m_\alpha} +  
	{m_\alpha \omega^2_\alpha \over 2}  
	\Bigl(x_\alpha + {c_\alpha \kappa_2  Q \over  m_\alpha \omega^2}  
	\Bigr)^2
	\Bigr]
\end{eqnarray}
which reduces to the non dissipative form by letting $c_\alpha=0$.
This is a convenient starting point for a weak coupling analysis also 
because a static shift in the equilibrium points 
of the oscillators has no effect even if it is large.

\end{multicols} 
\end{document}